\documentclass[aps,twocolumn]{revtex4}%
\usepackage{amsfonts}%
\usepackage{amsmath}%
\usepackage{amssymb}%
\usepackage{graphicx}
\providecommand{\U}[1]{\protect\rule{.1in}{.1in}}

\newcommand{\bel}[1]{\begin{equation}\label{#1}}
\newcommand{\bal}[1]{\begin{eqnarray}\label{#1}}
\newcommand{\ee}{\end{equation}}
\newcommand{\ea}{\end{eqnarray}}
\newcommand{\equ}[1]{~Eq.(\ref{#1})}

\newcommand{\bmath}{\begin{displaymath}}
\newcommand{\emath}{\end{displaymath}}
\newcommand{\bite}{\begin{itemize}}
\newcommand{\eite}{\end{itemize}}

\newcommand{\drop}[1]{}

\begin{document}
\preprint{HEP/123-qed}
\title[Casimir Geometrical Weakening]{Weak and Repulsive Casimir Force in Piston Geometries}
\author{Martin Schaden}
\affiliation{Department of Physics, Rutgers University, 101 Warren Street, Newark NJ 07102}
\author{Liviu Mateescu}
\affiliation{Department of Physics, New Jersey Institute of Technology, 323 King Blvd. Newark NJ 07102}
 \pacs{PACS number}

\begin{abstract}

We study the Casimir force in piston-like geometries semiclassically. The force on the piston is finite and
physical, but to leading semiclassical approximation depends strongly on the shape of the surrounding cavity.
Whereas this force is attractive for pistons in a parallelepiped with flat cylinder head, for which the
semiclassical approximation by periodic orbits is exact, this approximation to the force on the piston vanishes
for a semi-cylindrical head and becomes repulsive for a cylinder of circular cross section with a hemispherical
head. In leading semiclassical approximation the sign of the force is related to the generalized Maslov index of
short periodic orbits between the piston and its casing.

\end{abstract}





\startpage{101}
\endpage{120}
\maketitle

\section{Introduction}

Quite contrary to intuition derived from the Casimir force between two conducting plates\cite{Casimir},
Boyer\cite{Boyer1} found that the zero-point contribution to the surface tension of a perfectly conducting
spherical shell is negative. Until recently\cite{Schaden1}, there was no qualitative explanation of this result.
However, the finite negative surface tension of a metallic spherical shell cannot by itself be measured. Any
change in cavity radius necessarily involves material properties in a non-trivial way. The negative Casimir
tension of a spherical shell in this sense is a mathematical result without physical consequences. It was later
found that Casimir self-energies of many closed cavities are in fact plagued by divergences that cannot be
removed without appealing to material properties of the cavity walls\cite{Deutsch1}.

The Casimir force between distinct bodies on the other hand is in principle observable and must be finite. For
some simple geometries this force between uncharged conductors has now been measured quite accurately
\cite{Experiments}. Experimentally as well as theoretically the force between conductors is attractive in all
cases studied so far. A theorem by Kenneth and Klich \cite{Klich} and its recent generalization by Bachas
\cite{Bachas} states that the electromagnetic interaction between any mirror-pair of distinct (charge-conjugate)
bodies is attractive. This theorem in particular implies that contrary to previous
suggestions\cite{Lamoreaux,Elizalde}, the force between two half-spheres is attractive\cite{Klich}. The
attractive Casimir force between polarizable atoms furthermore suggests that the force might be attractive for
\emph{any} geometry of conductors. Such considerations, as well as failed attempts
\cite{Lamoreaux,Elizalde,Schmidt} at finding geometries with repulsion, could give the impression that repulsive
Casimir forces between distinct bodies arise only for suitable (mixed) boundary
conditions\cite{Hushwater,Barton,Fulling}.

Although some of these arguments are very suggestive, neither the long list of examples nor the restrictive
theorems by Kenneth, Klich and Bachas\cite{Klich,Bachas} imply that the Casimir force is attractive between
\emph{any} two conductors. Intuition based on the Casimir-Polder\cite{Casimir2} force between atoms can be
misleading\cite{Kenneth1}. Polarizable atoms attract just as any distant conducting spheres would and as such do
not give the Casimir force for other geometries. [If two-body forces were all one had to consider, a metallic
spherical shell should, for instance, have positive surface tension.]

Although no general explanation for the sign of Casimir forces has been given so far, the semiclassical
approximation tends to correctly predict the sign of Casimir energies and even provides reasonably accurate
estimates of their magnitude whenever the leading contribution from periodic classical rays does not happen to
vanish. Semiclassically, the Casimir force is related to optical properties of the cavity. From the
semiclassical point of view, the Casimir energy of two parallel plates and of a spherical shell are as different
as optical properties of flat and curved mirrors\cite{Schaden2}. The positive Casimir energy of the spherical
shell in this approximation is due to the presence of caustic surfaces\cite{Schaden1}. These lead to a relative
phase lag of the contributions from classical periodic rays that ultimately determines the sign of the Casimir
energy.

However, the contribution of periodic rays to the Casimir force/energy does not always tell the whole and
sometimes does not even tell the main story. The Casimir energy/force may vanish in this approximation. A
cylindrical conducting shell in three spatial dimensions perhaps is the best known
example\cite{Schaden1,Stecher}, but geometries without any periodic orbits, such as the Casimir
pendulum\cite{Jaffe}, are among these as well. The Casimir force/energy and in these cases depends entirely on
contributions to the spectral density of higher semiclassical order. These include lower
dimensional-\cite{Brackbook} and diffractive-\cite{Brackbook,Keller,Schaden3} contributions to the spectral
density that are associated with the presence of a boundary. Much of the elegance and predictive power of the
semiclassical approach is lost when periodic rays do not contribute and even the sign of the Casimir
force/energy may be difficult to estimate in this case. The semiclassical approach is more predictive when the
leading approximation due to classical periodic orbits does not vanish. In integrable systems the latter provide
a description of the spectral density that is dual to that of the cavity modes\cite{Brackbook}. The following
investigation assumes that corrections of higher semiclassical order will not dominate the leading estimate to
the Casimir force due to periodic orbits when the latter is appreciable. The corrections in particular should
not change the sign of the leading order estimate. Our semiclassical analysis of three piston geometries singles
out systems with interesting (because somewhat counterintuitive) semiclassical properties that in principle
could be verified numerically\cite{Gies} -- or perhaps even experimentally.

\begin{center}
\includegraphics[width=3.2truein]{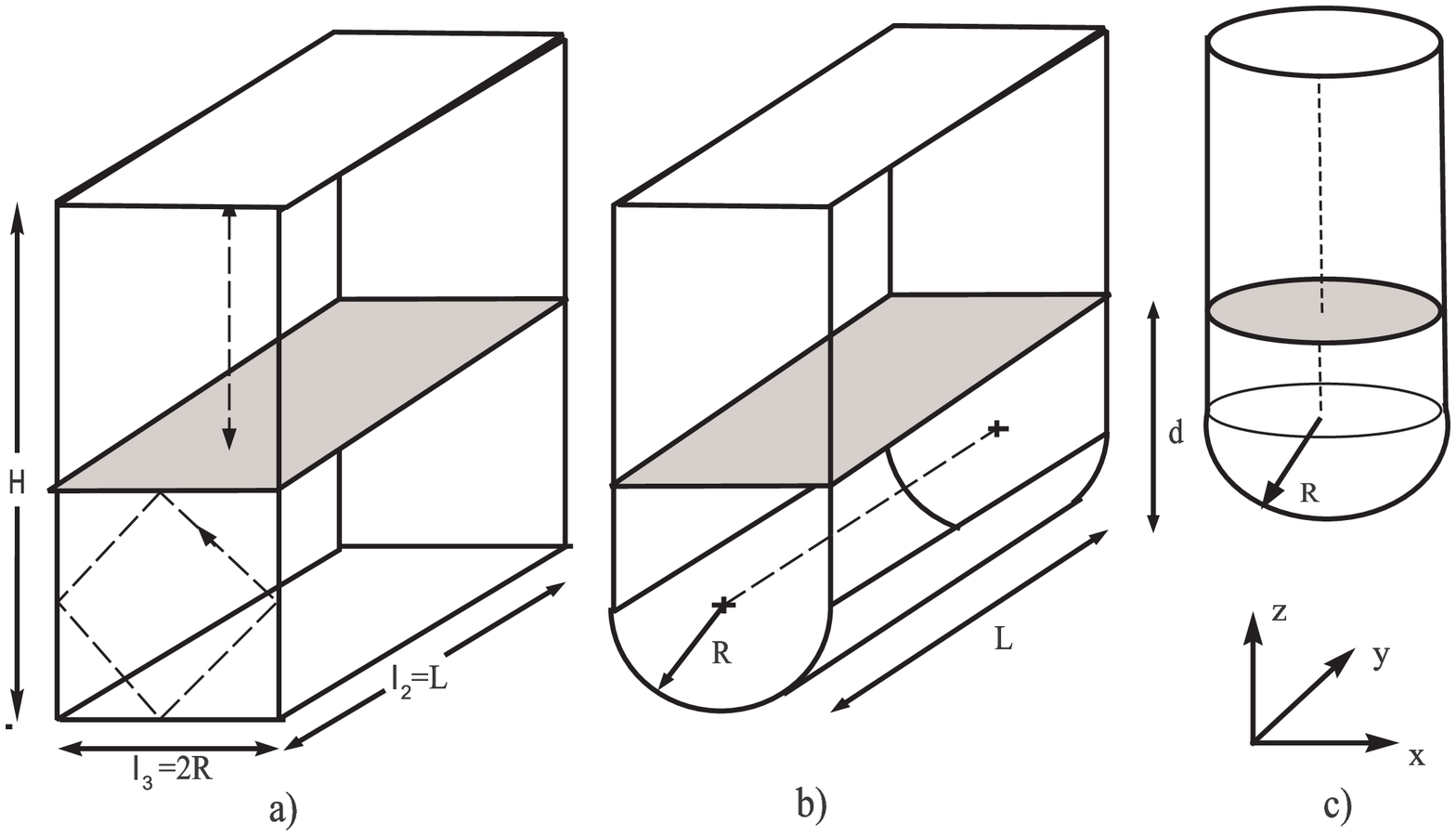}
\end{center}
\noindent{\small Fig.1 : Casimir piston geometries investigated here in the limit $H,L\gg R,d$. In leading
semiclassical approximation the electromagnetic Casimir force on the movable piston (shaded surface) is
attractive for cavity a), vanishes or changes sign for cavity b) and is repulsive at some $d>R$ for cavity c).
In Casimir pistons of type b) the direction of the force depends on whether Neumann, Dirichlet or metallic
boundary conditions are imposed. Two representative classical periodic rays that contribute to the force on the
piston are indicated as dashed lines in a).}

\section{Casimir Pistons}

\subsection{Parallelepiped with Flat Head}

The geometry of the Casimir piston shown in Fig.~1a) apparently was first used by Power\cite{Power} and clearly
demonstrates that the Casimir force between two flat and perfectly conducting mirrors is physical and does not
depend on calculational details such as cutoffs\cite{Svaiter}. We here refer to any cavity with an internal
dividing surface that is freely movable by Casimir forces as a "Casimir piston", a terminology coined
in\cite{Cavalcanti}. The force $F_p(d<H,l_2,l_3)$ on the dividing wall at a distance $d$ from one end of the
parallelepiped in general depends non-trivially on the dimensions of the parallelepiped and the position of the
piston. For $d\ll H, l_2,l_3$ the force approaches the one obtained by Casimir for two parallel conducting
plates\cite{Casimir},
\bel{Cas1}
F_p(d\ll l_2,l_3\ll H)\sim -\frac{\hbar c \pi^2 l_2 l_3}{240 d^4}\ .
\ee
Of greater interest to our investigation is that the force on the piston may be \emph{exactly} computed
semiclassically\cite{Brackbook} for any position of the wall and any dimension of the parallelepiped. The force
gives the dependence of a suitably subtracted zero-point energy on the height $d$ of the piston,
\bal{Power1}
F_p(d,H,l_2,l_3)&=&-\frac{\partial}{\partial d}\tilde{\cal E}_p(d,H,l_2,l_3)\\
\tilde{\cal E}_p(d,H,l_2,l_3)&=&\left[{\cal E}_p(d,l_2,l_3)+\right.\nonumber\\
&&\left. +{\cal E}_p(H-d,l_2,l_3)-2 {\cal E}_p(H/2,l_2,l_3)\right].\nonumber
\ea
Here ${\cal E}_p(l_1,l_2,l_3)$ is the (formal) zero-point energy of a parallelepiped with dimensions $l_1\times
l_2\times l_3$. Note that the (infinite, but $d$-independent) vacuum-energy of the Casimir piston at $d=H/2$ has
been subtracted. Svaiter emphasizes that the subtracted zero point energy in\equ{Power1} is finite and does not
depend on the cutoff procedure. These general considerations are quite independent of the nature of the walls
and of the exterior to the system and apply equally well to pistons and walls of finite thickness and finite
conductivity.

The force on the piston can be computed semiclassically for idealized boundary conditions, essentially because
the duality transformation from cavity modes to classical periodic paths (using Poisson's formula) can be
performed {\it exactly} \cite{Brackbook}. Contributions to the spectral density proportional to the total volume
of the parallelepiped, its total surface area and (total) edge length as well as contributions that reflect
topological features of this geometry, such as the number and type of corners\cite{Balian}, do not depend on $d$
and do not contribute to the subtracted Casimir energy in\equ{Power1}. The force on the piston is entirely due
to families of periodic classical rays that touch the piston as well as the enclosing parallelepiped. Their
contribution to the Casimir energy\emph{exactly} matches the field theoretic result\cite{Lukosz, Wolfram} for
this system. Some representative periodic rays in the bulk of the cavity are shown in Fig.~1a). All rays are of
finite length and the semiclassical expression for the (subtracted) Casimir energy is inherently finite. For
$l_2=L\gg l_3=2R$ it is readily seen that the plate is attracted\cite{Jaffe2} toward the nearer end of the
parallelepiped in the electromagnetic case\cite{Lukosz, Wolfram}. Boundary corrections\cite{Brackbook} due to
periodic rays that lie within the surfaces of the parallelepiped cancel for the electromagnetic
field\cite{Lukosz,Wolfram,Balian,Schaden4} and the contribution to the $d$-dependent part of the Casimir energy
from edges may be ignored\footnote{The Casimir force on the piston of Fig.~1a in fact is always
attractive\cite{Jaffe2} as $H\rightarrow\infty$, but sub-leading repulsive contributions from periodic boundary
rays that are entirely within the edges and surfaces complicate the analysis for general $l_2,l_3$.} for $l_2\gg
l_3$. Any family of periodic rays in the bulk of the cavity contributes negatively to the Casimir energy because
its Maslov index vanishes\cite{Schaden4} -- there are no caustics or focal points and the number of reflections
is even. Because the (negative) contribution to the Casimir energy of longer rays decreases in magnitude, the
piston is attracted toward the closer end of the cavity in this case.

This example of a parallelepiped-piston already hints at the possibility that the Casimir force could be
repulsive if the Maslov index of dominant (preferably all) classes of periodic rays would not vanish. The merit
of this conjecture is more apparent when the perfectly conducting piston is replaced by a perfectly permeable
one. Hushwater\cite{Hushwater} showed that the Casimir force between a perfectly conducting and a perfectly
permeable plate is repulsive and Fulling\cite{Fulling} recently extended these considerations to the piston
geometry. The change in sign of the Casimir force is readily explained semiclassically\cite{Schaden0} by a
change in the Maslov index of the relevant periodic orbits. Since the reflection coefficients of a classical ray
are of opposite sign on the permeable piston and the conducting parallelepiped, the Maslov index of a class of
periodic rays that reflect $n$ times off the piston is $2n$ modulo $4$. It corresponds to an overall phase lag
of $n\pi$. Periodic rays that reflect an odd number of times off the piston thus contribute positively to the
Casimir energy. Since rays with $n=1$ reflections off the piston are by far the shortest relevant periodic
orbits, their contribution dominates the Casimir energy and the force on a perfectly permeable piston surrounded
by a perfectly conducting parallelepiped is repulsive\cite{Schaden0}.

A highly permeable plate may be difficult to realize but this simple example links the sign of the semiclassical
estimate of the Casimir force to the Maslov index of the dominant (usually the shortest) classical periodic
rays\cite{Schaden4}. We now consider Casimir pistons for which the Maslov index of \emph{all} classes of
periodic rays differs from that in the parallelepiped geometry. The change in Maslov index in this case is of
geometrical (optical) origin rather than due to a change in boundary conditions. We will only consider Neumann,
Dirichlet and perfectly conducting surfaces.

\subsection{Parallelepiped with Semi-Cylindrical Head}
We first replace the flat head of the parallelepiped by a half-cylinder of radius $R$ as shown in Fig.~1b). The
rectangular cross section of this Casimir piston remains $2R\times L$. To simplify the analysis, we again study
the limit of large $L\gg R$.

In this limit we have a translational symmetry (chosen along the y-axis). Volume, surface and topological parts
of the spectral density cancel as before and we are again led to only consider contributions from classical
periodic rays of finite length that \emph{depend} on the height $d\ge R$ of the piston above the half-pipe. As
for a rectangular cavity, (classes of) periodic orbits that do not reflect off the piston contribute to the
spectral density and Casimir energy, but not to the Casimir \emph{force} on the piston.

The translational symmetry in $y$-direction essentially reduces our problem to the 2-dimensional one of
Fig.~2a). The spectral density of this 2-dimensional cavity decomposes into that of half a stadium and that of a
rectangle. The stadium billiard is a classically chaotic system that has been studied intensely and for general
$d>R$ the spectral density fluctuations appear to be best described by random matrix theory\cite{Hessel}. At
$d=R$ the stadium is a half-disc and most periodic orbits are degenerate. As shown in Figs.~2b) and c), the
degenerate periodic orbits of the half-cylinder (half-disc) are in direct correspondence with those of the
cylinder (disc). The latter may be identified by a pair of integers $(n,m)$, where $n$ is the number of
reflections off the cylinder and $m$ denotes the number of turns of the periodic orbit around the cylinder axis.
Evidently $2\leq 2 m\leq n$ and only classes of periodic orbits with $n=2m$ pass through the axis of the
cylinder. A half-cylinder can be viewed as a cylinder where points reflected about a plane through the cylinder
axis have been identified. All periodic orbits of the cylinder thus correspond to periodic orbits of the
half-cylinder, but the half-cylinder has some additional ones. Just as for the full cylinder, a class of
periodic orbits of the half-cylinder is identified by just two numbers: the integer number ($n$) of reflections
off the half-pipe and the number of times ($m$) the periodic orbit cycles between the two quadrants of the
half-cylinder. This classification fails only for two types of periodic orbits -- the up-down orbits on the
radian dividing the two quadrants of the half-cylinder and periodic orbits that touch an edge of the
half-cylinder. The latter are limiting cases within a family of periodic orbits that generally do not touch
either corner -- they therefore belong to these classes. However, the up-down periodic orbits of the
half-cylinder with an \emph{odd} number of reflections off the half-pipe have no analog among those of the full
cylinder. We will associate these, in the half-disc, isolated periodic orbits with a pair $(n,n/2)$ where $n$ is
an odd integer. This assignment is consistent with the previous one insofar as any \emph{even} number of
repetitions of the primitive $(1,1/2)$ orbit belongs to the degenerate class $(2m,m)$ with integer $m$, whereas
an \emph{odd} number of repetitions of this orbit results in an up-down orbit that has no analog in the full
cylinder.
\begin{center}
\includegraphics[width=3.2truein]{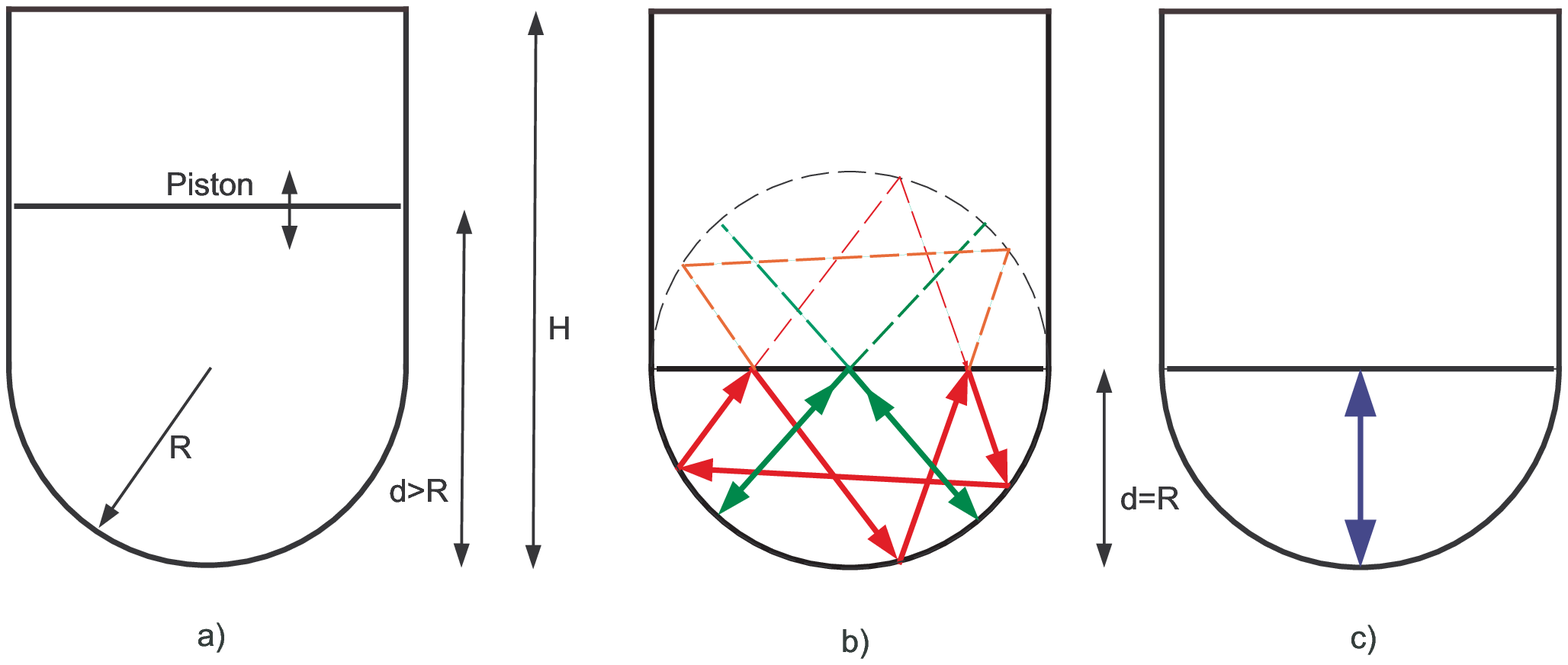}
\end{center}
\noindent{\small Fig.~2: a) Two-dimensional cavity with semi-circular head and a piston at $d>R$. b)
representative primitive $(3,1)$ and $(2,1)$ orbits when $d=R$. The corresponding reflected orbits in the
$(3,1)$ and $(2,1)$ classes of the full cylinder are shown as dashed extensions. Note that two representatives
of a class of periodic orbits of the full disc correspond to the same representative in the half-disc. c) The
primitive $(1,1/2)$ orbit of the half-disc.}

Note that an $(n,m)$-family of periodic orbits in the half-cylinder to integer $m$ covers the $(n,m)$-family of
the full cylinder exactly twice. The contribution from classical periodic rays to the spectral density of a
half-cylinder thus may be written,
\bel{semirhocyl}
\tilde\rho_{\rm cyl./2}(E)=\frac{1}{2}\tilde\rho_{\rm cyl.}(E)+\sum_{k=1}^\infty \tilde\rho^{\rm
halfcyl.}_{(2k-1,k-{\frac1 2})}(E)\ .
\ee
The first term in \equ{semirhocyl} is just half the corresponding spectral density of a cylinder due to classes
of periodic orbits $(n,m)$ to integer $m$. The correction is from up-down orbits $n=2m$ to odd $n$. The
contribution to the zero-point energy of a half-cylinder from periodic orbits may be similarly decomposed,
\bal{Ehalfcyl}
{\cal E}_{\rm cyl./2} &=& \frac{1}{2}\int_0^\infty\hspace{-1em}E  \tilde\rho_{\rm cyl./2}(E)dE\nonumber\\
&=& \frac{1}{2} {\cal E}_{\rm cyl.}+ {\cal E}_{\rm ud-cyl.}\ ,
\ea
with
\bel{Eupdown}
{\cal E}_{\rm ud-cyl.}=\frac{1}{2} \sum_{k=1}^\infty \int_0^\infty\hspace{-1em}E \tilde\rho^{\rm
halfcyl.}_{(2k-1,k-{\frac1 2})} dE
\ee
 The semiclassical contribution from periodic rays to the Casimir energy of a perfectly conducting cylinder
vanishes\cite{Stecher,Schaden4}. We here show that the semi-classical Casimir energy of a half-cylinder vanishes
as well. We use the extension of Gutzwiller's trace formula \cite{Gutzwiller} developed by Creagh and
Littlejohn\cite{Brackbook,Creagh1,Creagh2} to include continuous symmetries. The translational symmetry is a
particularly simple Abelian symmetry generated by the momentum $p_y$ along the $y$-axis. For a periodic orbit
$\Gamma$ of length $l_\Gamma$ in the symmetry-reduced space, the Jacobian is
\bel{Jacobian} J_\Gamma={\rm det} \frac{\partial y}{\partial
p_y}=\frac{l_\Gamma}{p}=\frac{l_\Gamma c}{E}
\ee
The volume of the translation group is just the total length $V_\Gamma=L\gg R$ of the half-cylinder. The
primitive period of an up-down ray is $T_\Gamma=2 R/c$ and its classical action is simply $S_\Gamma=\oint {\bf
p\cdot dx}=E l_\Gamma/c$. The semiclassical expression for the spectral density derived in\cite{Creagh1} (with
$f=1$ constants of motion) for our special case is,
\bal{spectraldens} \tilde\rho_\Gamma(E) &=&\frac{T_\Gamma
V_\Gamma}{\pi\hbar}\frac{\cos\left(S_\Gamma/\hbar-\sigma_\Gamma
\frac{\pi}{2}-\frac{\pi}{4}\right)}{|2\pi\hbar J_\Gamma \det(M_\Gamma-1)|^{1/2}}\\
&=& \frac{L R}{\pi\hbar cl_\Gamma}\left(\frac{E l_\Gamma}{2\pi\hbar
c}\right)^{1/2}\cos\left(\frac{E l_\Gamma}{\hbar c}-\sigma_\Gamma
\frac{\pi}{2}-\frac{\pi}{4}\right)\ ,\nonumber
\ea
where we have used that the reduced stability matrix $M_\Gamma$ of an up-down ray with an \emph{odd} number of
reflections off the half-pipe has two eigenvalues equal to $-1$ (the orbit is inverse parabolic) and thus
$\sqrt{|\det(M_\Gamma-1)|}=2$. [The stability matrix of an up-down ray with an \emph{even} number of reflections
off the half-pipe has eigenvalues $\lambda=1$ and, as we have seen, represents a class of degenerate orbits of
the cylinder.] The dependence of the spectral density of up-down orbits on $E$ is explicit in\equ{spectraldens}
and one can perform the integration in \equ{Eupdown} to obtain,
\bel{Eupdown1} {\cal E}_{\rm ud-cyl.}=\frac{3 L R\hbar c}{8\pi
\sqrt{2}}\sum_{\Gamma\in \{(2k-1,k-1/2)\}}\frac{-\cos\left(\sigma_\Gamma\frac{\pi}{2}\right)}{l^3_\Gamma}\ .
\ee
This is a particular example of the fact that the sign of the contribution of an isolated classical period orbit
is determined by its generalized Maslov index\cite{Schaden4}. The Maslov index of an isolated orbit is the sum
of two integers\cite{Gutzwiller}, $\sigma_\Gamma=\mu_\Gamma+\nu_\Gamma$. $\nu_\Gamma$ gives the number of
independent transverse directions in which a movement of the starting/endpoint reduces the classical action of
the closed orbit and is the number of negative eigenvalues of Gutzwiller's stability
matrix\cite{Brackbook,Gutzwiller}. The following mechanical analogy often makes $\nu_\Gamma$ "obvious" for
billiard systems without much calculation. Since the classical action of any billiard is proportional to the
length of the periodic orbit, one simply thinks of the trajectory as an elastic band. The index $\nu_\Gamma$ at
any point on the orbit then is the number of independent transverse directions in which a (slight) pull would
cause the orbit to slip and not return to its original position upon release. From this mechanical analogy it is
quite clear that $(2k-1,k-1/2)$-rays have a stability index $\nu_{(2k-1,k-1/2)}=1$ that does not depend on the
choice of starting point on the orbit. The integer $\mu_\Gamma$ is determined from the number and order of
conjugate points and the boundary conditions. The number of reflections on an up-down orbit is always even. For
Dirichlet or Neumann conditions on \emph{both}, the piston \emph{and} the half-pipe, reflections thus contribute
an irrelevant multiple of $2\pi$ to the overall phase. Only the number of conjugate points on the orbit matters.
$\sigma_\Gamma=\mu_\Gamma+\nu_\Gamma$ is a topological characteristic of the periodic orbit\cite{Creagh3} that
does not depend on the starting point and we have seen that $\nu=1$ for up-down trajectories also does not
depend on this choice. One therefore can obtain the number of conjugate points using any starting point on the
orbit. Since approximately paraxial rays remain approximately paraxial for many cycles and cross the y-axis near
the focal point at $z=R/2$ (they cross in the interval $[(3-\sqrt{5})R/2,R/2]$), there are $(2k-2)$ conjugate
points on an $(2k-1,k-1/2)$-ray. All conjugate points are of first order and the generalized Maslov index of
up-down rays (with Dirichlet or Neumann boundary conditions) therefore is odd,
\bal{Maslov}
\sigma_{(2k-1,k-1/2)}&=&\mu_{(2k-1,k-1/2)}+\nu_{(2k-1,k-1/2)}\nonumber\\
&=&2k-1 \mod 4\ .
\ea
${\cal E}_{ud}$ defined in \equ{Eupdown1} thus vanishes for Dirichlet (D) as well as Neumann (N) boundary
conditions. The results for a cylinder\cite{Brackbook} then give,
\bal{halfcyl}
{\cal E}^{\left(D\atop N\right)}_{cyl./2}&=&\frac{1}{2}{\cal E}^{\left(D\atop N\right)}_{cyl.}\\
&=&(\pm)\frac{L\hbar c}{32\pi R^2}\sum_{m=1}^\infty\sum_{k=m}^\infty \frac{(-1)^k}{
(2k+1)^4\sin^2\left(\frac{m\pi }{2k+1}\right)}\ \nonumber\\
&=&(\mp) 0.0001209\dots\frac{L\hbar c}{R^2}\nonumber\ ,
\ea
where the upper sign corresponds to Dirichlet (D) and the lower to Neumann (N) boundary conditions . The
electromagnetic Casimir energy of a cylinder (or any cavity with a one-dimensional translational symmetry) is
just the sum of the contributions from two decoupled scalar fields satisfying Dirichlet and Neumann boundary
conditions. From\equ{halfcyl} it is apparent that the contribution due to periodic rays to the electromagnetic
Casimir (self-)energies of a cylinder and a half-cylinder both vanish,
\bel{CasEMcyl}
{\cal E}^{EM}_{cyl./2}=\frac{1}{2}{\cal E}^{EM}_{cyl.}=0\ .
\ee
By contrast, the field theoretic result for the Casimir energy of an infinitesimally thin but perfectly metallic
half-cylinder diverges\cite{Nesterenko01} due to arbitrary short closed paths near the sharp edges. The field
theoretic electromagnetic Casimir self-energy of a cylinder with metallic boundaries also does not vanish and
has the finite negative value ${\cal E}^{EM}_{\rm cyl.}({\rm field theory})=-0.1356\dots L\hbar c/R^2$.  Before
dismissing the semi-classical result of\equ{CasEMcyl}, note that the field theoretic calculation of the Casimir
self-energy of a cylinder and half-cylinder includes contributions that are irrelevant to the \emph{force} on
the piston of Fig.~2b. Among these are contributions from exterior modes\cite{DRM81} as well as ultraviolet
divergent contributions to the self-energy of a half-cylinder due to its sharp edges\cite{Balian}. The latter do
not change when the piston is moved and do not contribute to the force on it. The semiclassical contribution to
the Casimir energy due to classical periodic orbits that we have calculated on the other hand certainly
\emph{depends} on the position of the piston. As for the rectangular piston of Fig.~1a), no net force results
from periodic orbits that \emph{do not} touch the piston \emph{and} the surrounding cavity. The contribution to
the Casimir energy from the corresponding periodic orbits of the $2 R\times H\times L$ dimensional
parallelepiped on the other side of the piston is finite and negative, but readily seen to decrease with
increasing height $H$. For dimensional reasons this contribution to the vacuum energy of the cylinder vanishes
as $\hbar c L R/H^3$. For $d\gg 2 R$, periodic orbits that touch the piston and the half-pipe at least have
length $2 d$ and their contribution to the Casimir energy is proportional to $\hbar c L R/d^3$ on dimensional
grounds. For $H\sim\infty$ we conclude that the electromagnetic Casimir energy of the cavity in Fig.~1b) due to
periodic orbits vanishes at two positions of the piston,
\bel{vanEM}
{\cal E}^{EM}_{\rm 1b)}(d=R)={\cal E}^{EM}_{\rm 1b)}(d=H/2\sim\infty)=0\ .
\ee
Note that this equality holds only for the (semiclassical) contribution from periodic orbits. If this were the
only $d$-dependent contribution, \equ{vanEM} would imply that the Casimir force on a rectangular piston within a
parallelepiped of dimensions $2R\ll L,H$ capped by a half-cylindrical head either vanishes for any $d>R$, or is
repulsive for some $d>R$. Diffractive corrections we have not computed may alter this conclusion, but it
nevertheless is quite striking that the semiclassical estimate of the Casimir energy due to periodic orbits
differs drastically when the half-cylinder of Fig.~1b) replaces the flat piston head of Fig.~1a). The situation
is slightly more dramatic for a massless scalar field satisfying Neumann boundary conditions on all surfaces.
The previous considerations for $L,H\gg 2R$ together with \equ{halfcyl} in this case lead to the conclusion
that,
\bel{repN}
{\cal E}^{N}_{\rm 1b}(d=R)\sim 0.0001209\frac{L\hbar c}{R^2}>{\cal E}^{N}_{\rm 1b}(d= H/2\sim\infty)\ ,
\ee
and imply that the leading semiclassical contribution to the Casimir force on the piston is repulsive for some
$d>R$. Diffractive corrections in this case would have to overwhelm the (numerically small) leading
semiclassical contribution from periodic orbits for the Casimir force on the piston to be attractive at all
$d>R$.

\subsection{Cylinder with Hemispherical Head}
Massless scalars satisfying Neumann boundary conditions are not readily available and our intuition is mainly
based on the attractive nature of the Casimir force between polarizable atoms. The previous example of a
semiclassical Casimir force that is repulsive for geometrical reasons thus is only of academic interest.
However, it illustrates that the leading semiclassical contribution to the Casimir force may change sign for
cavities with the appropriate optical properties. The metallic cylindrical cavity with a piston and a
hemispherical head of Fig.~1c) still may be difficult to realize, but has the distinct advantage that in leading
semiclassical approximation the force on the piston is repulsive also in the electromagnetic case.

The argument closely follows that for the half-cylindrical head, with just two (crucial) modifications. We again
consider only the contribution of periodic orbits that reflect off the piston and the enclosing cylinder. For
$d=R$ their contribution to the semiclassical Casimir energy of the cavity is that for a half-sphere. The
remaining contribution to the Casimir force due to periodic orbits in the upper cylinder becomes vanishingly
small with increasing length $\infty\sim H\gg 2 R$ of the cylinder. [As before, the force on the piston does
\emph{not} arise due to periodic orbits in planes perpendicular to the cylinder axis. When the piston is
repositioned, the phase space of transverse orbits lost on one side is regained on the other side of the piston.
Although these orbits give the dominant contribution to the oscillatory spectral density of a long cylinder,
they do not contribute to the net force on the piston.] For $d=R$ we again may decompose the Casimir energy of
the half-sphere into half that of the sphere and that due to (in this case) \emph{isolated} up-down
$\{(2k-1,k-1/2),k=1,2,\dots\}$-orbits along the $z$-axis with an odd number of reflections off the hemisphere,
\bel{Ehalfsphere}
{\cal E}_{\rm sp./2} =\frac{1}{2} {\cal E}_{\rm sp.}+ {\cal E}_{\rm ud-sp.}\ ,
\ee
with
\bel{Eupdownsp}
{\cal E}_{\rm ud-sp.}=\frac{1}{2} \sum_{k=1}^\infty \int_0^\infty\hspace{-1em}E \,\tilde\rho^{\rm
halfsp.}_{(2k-1,k-{\frac 1 2})} dE \ .
\ee

Contrary to the cylindrical case, periodic classical orbits contribute \emph{positively} and give about 99\% of
the field theoretic Casimir self-energy of an infinitesimally thin but perfectly conducting spherical
shell\cite{Boyer1,Schaden1} . [The rather small difference to the field theoretic self-energy perhaps is due to
the fact that the Casimir energy of the boundary of a 3-dimensional ball, that of a two-sphere, vanishes.]

The up-down orbits in this case are isolated and the corresponding semiclassical contribution to the spectral
density of a half-sphere is given by Gutzwiller's trace formula\cite{Gutzwiller}, \
\bel{Gutzwiller}
\tilde\rho_\Gamma=\frac{1}{\hbar\pi} \frac{T_\Gamma\cos\left(S_\Gamma/\hbar-\sigma_\Gamma
\frac{\pi}{2}\right)}{|\det(M_\Gamma-1)|^{1/2}}  \ .
\ee
As before, the primitive period of any up-down ray is $T_{\rm ud}=2 R/c$, but the stability matrix $M_\Gamma$
here is 4-dimensional. For $(2k-1,k-1/2)$-rays all four eigenvalues equal $-1$ and
$|\det(M_{(2k-1,k-1/2)}-1)|^{1/2}=4$. The integral over the energy $E$ of\equ{Eupdownsp} can again be performed
with the result,
\bel{Eupdownsp1}
{\cal E}^{\left(D\atop N\right)}_{\rm ud-sp.}=-\frac{\hbar c R}{4\pi}\sum_{\Gamma\in \{(2k-1,k-{\frac 1
2})\}}\frac{\cos(\sigma_\Gamma {\frac \pi 2})}{l^2_\Gamma}\ ,
\ee
irrespective of whether Neumann or Dirichlet boundary conditions hold. The other difference to the cylindrical
case is in the generalized Maslov index $\sigma_\Gamma$ of the up-down rays. It is computed as before, but there
now are \emph{two} independent unstable directions and $\nu=2$ for all up-down orbits. In addition, the
conjugate points of an $(2k-1,k-1/2)$-orbit are of second order for the hemisphere and each of the $2k-2$
conjugate points thus increase $\mu$ by 2. The overall generalized Maslov index of an $(2k-1,k-1/2)$-orbit thus
is,
\bel{Maslovsphalf}
\sigma_{(2k-1,k-1/2)}=\mu_{(2k-1,k-1/2)}+\nu_{(2k-1,k-1/2)}= 2\mod 4 \ .
\ee
Up-down orbits therefore contribute positively to the Casimir energy of a half-sphere,
\bal{Eupdownsp2}
{\cal E}^{\left(D\atop N\right)}_{\rm ud-sp.}&=&\frac{\hbar c}{16\pi R}\sum_{k=1}^\infty (2k-1)^{-2}\nonumber\\
&=&\frac{\hbar c\pi}{128 R}\sim 0.02454\dots \frac{\hbar c}{R}\ .
\ea
Since the semiclassical Casimir energy due to periodic rays of a spherical cavity is\cite{Schaden1},
\bel{Cassphere}
{\cal E}^{\left(D\atop N\right)}_{\rm sp.} =\frac{\hbar c}{32\pi R}\left[\sum_{k=1}^\infty \frac{1}{k^4}
+\sum_{k=2}^\infty\frac{15\pi \sqrt{2}}{16 k^4} \sum_{m=1}^{k-1}\frac{\cos(\frac{m\pi}{2k})}{
\sin^2(\frac{m\pi}{2k})}\right]\ ,
\ee
that of the half-sphere becomes,
\bal{Cashalfsphere}
{\cal E}^{\left(D\atop N\right)}_{\rm sp./2} &=&\frac{\hbar c\pi}{128 R}\left[1+\frac{\pi^2}{45}
+\sum_{k=2}^\infty\frac{15\sqrt{2}}{8\pi k^4} \sum_{m=1}^{k-1}\frac{\cos(\frac{m\pi}{2k})}{
\sin^2(\frac{m\pi}{2k})}\right]\nonumber\\
&=&\frac{1}{2}{\cal E}^{EM}_{\rm sp./2} \sim 0.03621...\frac{\hbar c}{R}\ .
\ea
The contribution of periodic orbits to the Casimir energy of a half-sphere does not depend on whether Dirichlet
or Neumann conditions are imposed. To leading semiclassical order, the electromagnetic spectral density due to
periodic rays for a perfectly conducting cavity again is just that of two massless scalars satisfying Neumann
and Dirichlet boundary conditions respectively. The Casimir energy from periodic rays entirely within the
circular edge where the piston joins the cavity does not depend on $d$ and therefore does not contribute to the
force. Note that up-down orbits give about $2/3$ of the total. This is consistent with the fact that the
primitive up-down ray is the shortest and all others have at least twice its length. Assuming that the energy
scales inversely with the length of the shortest primitive orbit when the piston is (slightly) moved, gives an
order-of-magnitude estimate of the repulsive force on the piston,
\bel{repforce}
F^{EM}_{1c}(d=R)\sim 0.07 \frac{\hbar c}{R^2}\sim \frac{2\times 10^3 pN}{(R{\rm~in~}nm)^2}\ .
\ee
The force is extremely weak and can barely support the weight(!) of a 75nm diameter and a few nanometer thick
graphite "piston".

\section{Conclusions}

The Casimir force on the pistons of the three geometries shown in Fig.~1 is finite for any (non-singular)
boundary condition. It in particular does not diverge for material surfaces of finite thickness and arbitrary
reflection coefficients. Contrary to the Casimir self-energy of the cavity itself, the force on the piston in
particular is finite for idealized Dirichlet, Neumann and perfectly conducting boundary conditions in all cases.
Movement of the piston does not change the ultra-violet divergence\cite{Deutsch1} of the zero-point energy of
the cavity and such contributions to the self-energy are subtracted by referring to a standard configuration of
the Casimir piston. One thus can study physical consequences of adiabatic changes in geometry in these simple
geometries and avoid conclusions that depend on mathematical properties of idealized boundaries and/or
regularization and subtraction schemes.

The rectangular piston of Fig.~1a) was discussed by Power\cite{Power} in connection with the original Casimir
force between conducting plates. The zero-point energy due to interior modes of a parallelepiped can be computed
exactly\cite{Lukosz,Wolfram} and is \emph{reproduced} by the semiclassical contribution due to periodic rays.
Using Poisson's formula, the duality transformation can be explicitly performed in this case\cite{Brackbook}.
However, only classical periodic trajectories in either parallelepiped that depend on the position of the piston
contribute to the force on it. These periodic rays have finite length and give a finite contribution to the
spectral density. For Neumann, Dirichlet and metallic boundary conditions, the internal piston is
attracted\cite{Jaffe2} to the nearer end of the cavity when $H\gg L, 2R$. The force in the parallelepiped-system
is \emph{entirely} given semiclassically by classes of periodic rays that reflect off the piston.

For the Casimir pistons shown in Figs.~1b)~and~1c) the duality transformation cannot be performed in closed form
and the semiclassical approximation no longer is exact. However, this approximation presumably does not suddenly
fail \emph{qualitatively} when the flat cylinder head of Fig.~1a) is replaced by the half-cylindrical one of
Fig.~1b) or the piston geometry is that of Fig.~1c). The length and number of classical periodic rays is similar
for all three cases, but the character of the orbits changes. Although the integrable parallelepiped is compared
to (for general $d>R$) chaotic systems, semiclassical periodic orbit theory should give reasonable estimates of
the spectral density, and in particular the dependence on cavity dimensions of its lowest moments.

Assuming that the semiclassical approximation by periodic orbits remains qualitatively correct, we have seen how
the Casimir force on a piston may be diminished and may even change sign for cavities with the appropriate
optical properties. For a repulsive Casimir force, the dominant (short and relevant) periodic orbits must have a
Maslov index that is an odd multiple of two. This can be achieved by either changing reflection properties of
the boundary or by  changes in geometry that introduce focal points or caustics. Since highly permeable surfaces
are difficult to realize, the latter option could be more practical. However, the force on nano-scale Casimir
pistons is extremely small, and it could prove difficult to measure it experimentally. Recent advances in the
world-line approach\cite{Gies} to Casimir effects potentially allow numerical studies of the Casimir pistons we
have here investigated semiclassically. The piston geometry of Fig.~1c) is repulsive even for Dirichlet boundary
conditions. The main numerical challenge appears to be the (accurate) subtraction of divergent contribution to
the Casimir energy that arise from arbitrarily short closed world lines that span the edge between the piston
and the cylinder, but it may be possible to directly compute the finite force. It may also be possible to study
Casimir pistons by field theoretic methods, perhaps by perturbing about the degenerate $d=R$-limit in which
these cavities reduce to composite systems with a high degree of symmetry. The field theoretic Casimir
self-energy of a half-cylinder diverges\cite{Nesterenko01}, but this divergence evidently is due to the sharp
edges and should not carry over to the force on the piston. Generalized Casimir pistons can help determine parts
of Casimir self-energies that are relevant for adiabatic changes of the geometry of a cavity. Such physical
Casimir energies do not depend on the contribution to the zero-point energy of exterior modes and are free of
ultraviolet divergences. They furthermore ought not to drastically depend on the particular type of idealized
boundary one might consider. Since the semiclassical contribution from periodic orbits incorporates all these
characteristics, this approximation might provide a useful guide and perhaps even a reasonable estimate to
physical Casimir forces.

\begin{acknowledgements}
This work is dedicated to the memory of Dr.~Larry Spruch who recognized that semiclassical analysis may provide
a qualitative and semi-quantitative understanding of Casimir effects. This work was supported by the National
Science Foundation with Grant No. 0555580.
\end{acknowledgements}




\end{document}